\documentclass[aps, prd, amsmath, twocolumn, preprintnumbers, superscriptaddress]{revtex4-1}
\usepackage{amssymb,esvect,amsmath,graphicx,latexsym,amsthm,slashed,eso-pic,hyperref}
\usepackage{xcolor}
\usepackage[final]{pdfpages}
\usepackage{dcolumn}
\newcolumntype{d}[1]{D{.}{.}{#1}}
\newcolumntype{C}{>{$}l<{$}}

\makeatletter
\AtBeginDocument{\let\LS@rot\@undefined}
\makeatother
  \DeclareMathOperator{\mev}{MeV}   \DeclareMathOperator{\fm}{fm} \DeclareMathOperator{\cm}{cm}     \DeclareMathOperator{\s}{s}    \DeclareMathOperator{\few}{few}  
     \newcommand{\cL}{{\cal L}} \newcommand{\cM}{{\cal M}}  \newcommand{\cO}{{\cal O}}

   \def\oL{\overline}
\newcommand{\pL}{\left(} \newcommand{\pR}{\right)} \newcommand{\bL}{\left[} \newcommand{\bR}{\right]} \newcommand{\cbL}{\left\{} \newcommand{\cbR}{\right\}} \newcommand{\mL}{\left|} \newcommand{\mR}{\right|}
\newcommand{\beq}{\begin{equation}} \newcommand{\eeq}{\end{equation}}
\newcommand{\bea}{\begin{eqnarray}} \newcommand{\eea}{\end{eqnarray}}
\newcommand{\alg}[1]{\begin{align} \begin{split} #1 \end{split}  \end{align}}
\newcommand{\vev}[1]{\langle {#1} \rangle}

\newcommand{\tenx}[1]{\times 10^{#1}}
\newcommand{\Eq}[1]{Eq.~(\ref{#1})} \newcommand{\Eqs}[2]{Eqs.~(\ref{#1}) and (\ref{#2})} \newcommand{\Eqm}[2]{Eqs.~(\ref{#1}) through (\ref{#2})}
\newcommand{\Sec}[1]{Sec.~\ref{#1}}  
\newcommand{\Fig}[1]{Fig.~\ref{#1}} 

\newcommand{\App}[1]{App.~\ref{#1}}

\begin{document}

\title{A Deeply Bound Dibaryon is Incompatible with Neutron Stars and Supernovae}
\author{Samuel D. McDermott}
\affiliation{Fermi National Accelerator Laboratory, Theoretical Astrophysics Group, Batavia, IL, USA}
\author{Sanjay Reddy}
\affiliation{Institute for Nuclear Theory, University of Washington, Seattle, WA, USA}
\author{Srimoyee Sen}
\affiliation{Institute for Nuclear Theory, University of Washington, Seattle, WA, USA}
\date{\today}

\begin{abstract}
We study the effect of a dibaryon, $S$, in the mass range $1860 \mev < m_S < 2054 \mev$, which is heavy enough not to disturb the stability of nuclei and light enough to possibly be cosmologically metastable. Such a deeply bound state can act as a baryon sink in regions of high baryon density and temperature. We find that the ambient conditions encountered inside a newly born neutron star are likely to sustain a sufficient population of hyperons to ensure that a population of $S$ dibaryons can equilibrate in less than a few seconds. This would be catastrophic for the stability of neutron stars and the observation of neutrino emission from the proto-neutron star of Supernova 1987A over $\sim \cO(10)\s$. A deeply bound dibaryon is therefore incompatible with the observed supernova explosion, unless the cross section for $S$ production is severely suppressed.
\end{abstract}

\preprint{FERMILAB-PUB-18-490-A~ ~INT-PUB-18-049 }

\maketitle

\section{Introduction}

The possibility that six light quarks form the QCD bound state $uuddss$, known as the $H$ dibaryon with binding energy $B_H \equiv 2 m_\Lambda - m_H \gtrsim 0$, has been considered for several decades~\cite{Jaffe:1976yi}. Direct searches from accelerator-based experiments have ruled out the possibility that such a state has weak decays that are easily detected \cite{Belz:1995nq, AlaviHarati:1999ds, Kim:2013vym, Tscheuschner, Adam:2015nca} or that such a state is more massive than approximately 2 GeV \cite{Badier:1986xz, Bernstein:1988ui}. The suggestion that a much more deeply bound state~\cite{Farrar:2003qy} called the $S$ sexaquark~\cite{Farrar:2017eqq,Farrar:2017ysn}, with $B_S \equiv 2 m_\Lambda - m_S \geq m_\Lambda - (m_p+m_e) = 176.9 \mev$ and which nontrivially avoids these observational bounds \cite{Zaharijas:2004jv, Farrar:2018hac}, deserves further scrutiny. Lattice studies will eventually be able to test the full spectrum of six-quark states and conclusively decide if such a state exists. Present studies support the existence of a weakly bound dibaryon with $B_H \sim \cO(10)\mev$~\cite{Beane:2010hg, Beane:2011zpa, Beane:2011iw, Beane:2012vq}, but the more tightly bound and thus stable or cosmologically metastable sexaquark with $B_S \sim \cO(\few \times 100) \mev$, cannot be ruled out at the current level of understanding of lattice systematics~\cite{SavageBeaneCommunication}.

In this work, we consider an $S$ that is light enough to be metastable but massive enough that it is not exothermically produced as a fusion product of two nucleons. This gives the constrained mass range $1860 \mev < m_S < m_\Lambda + m_p + m_e \simeq 2054 \mev$, which in turn implies
\beq \label{bs-range}
 176.9\mev < B_S < 361 \mev.
\eeq
Due to its electric neutrality and its (meta)stability, such a particle would be a candidate for the dark matter of the universe \cite{Farrar:2017ysn}. Such a state would avoid detection in underground direct detection experiments due to the overburden of earth, and may inefficiently deposit energy in the only relevant high-altitude direct detection search \cite{Mahdawi:2018euy}. The sexaquark would further have a small enough elastic scattering cross section to avoid present-day cosmological constraints from the power spectrum of the Cosmic Microwave Background Radiation \cite{Gluscevic:2017ywp} or from astrophysical gamma ray searches \cite{Hooper:2018bfw}.

The range of binding energies in \Eq{bs-range} ineluctably leads to the conclusion that the production of dibaryons from $\Lambda$ baryons is on-shell and exothermic, however. We study the implications of the production of such a deeply bound QCD state in hot proto-neutron stars. We conclude that observations are in grave tension with the hypothesis of a deeply bound $S$ unless the $S$ production cross section is highly suppressed.

\section{Baryons and Dibaryons in a Proto-Neutron Star}
Production and decay of the $S$ dibaryon is suppressed under ordinary conditions, because creating two units of strangeness requires a doubly weak process. However, the temperature and densities encountered in a proto-neutron star formed during a core-collapse supernova are large enough to produce a thermal population of hyperons through weak reactions~\cite{Pons:1998mm, Keil:1995hw}. Further, since temperatures of the order of tens of MeVs are sustained for a period of about 10 seconds -- a time scale set by neutrino diffusion from the proto-neutron star \cite{Burrows:1986me} -- we will demonstrate that reactions involving hyperons equilibrate the number density of the $S$ dibaryon except under the most extreme possible assumptions.
 
We begin by writing the coupled differential equations for the number density of different species of baryons. We include only the $N=n,p$ and $\Lambda$ states; charge conservation is implicit throughout. $\Lambda$'s can be produced either by the leptonic process $e^-+p\rightarrow \Lambda + \nu_e$, or by the non-leptonic process $NN  \rightarrow N\Lambda$ and $n \pi \rightarrow \Lambda$. Due to the high baryon density expected in the neutron star we shall ignore leptons for simplicity. The time evolution of the number density of each species $a$ is of the schematic form $\dot n_a = \text{(rate of $a$ production per unit volume)} - \text{(rate of $a$ disappearance per unit volume)}$. Because baryon number $\sf B$ is conserved, we expect that the rate of $N$ decay (production) is proportional to $n_N$ ($n_\Lambda$), and vice versa.
With this in mind, we write:
\begin{subequations}\begin{align} 
\dot n_N &= - n_N^2 \langle \sigma_{N N \to \Lambda N} v \rangle
- n_N n_\pi \langle \sigma_{N \pi \to \Lambda} v \rangle 
+ \nonumber \\
& ~~+ \frac{n_\Lambda}{\tilde \tau_\Lambda}  
+ n_\Lambda n_N  \langle \sigma_{N \Lambda \to N N} v \rangle
\label{diff-eqN}
\\
\dot n_\Lambda &= + n_N^2 \langle \sigma_{N N \to \Lambda N} v \rangle+ n_N n_\pi \langle \sigma_{N \pi \to \Lambda} v \rangle - \nonumber \\
&\label{diff-eqL}  ~~- \frac{n_\Lambda}{\tilde \tau_\Lambda}  
- n_\Lambda n_N  \langle \sigma_{N \Lambda \to N N} v \rangle -
\\ \nonumber &~~- 2 n_\Lambda^2 \langle \sigma_{ \Lambda \Lambda \to S X} v \rangle +  2 n_S n_X \langle \sigma_{ S X \to \Lambda \Lambda } v \rangle 
\\
\dot n_S &= + n_\Lambda^2 \langle \sigma_{ \Lambda \Lambda \to S X} v \rangle - n_S n_X \langle \sigma_{ S X \to \Lambda \Lambda } v \rangle \label{diff-eqS},
\end{align}\end{subequations}
where $\langle \sigma_i v \rangle$ indicates the thermally averaged cross section times velocity for the process $i$; we discuss the values of the various $\langle \sigma_i v \rangle$ in the ensuing sections. The particle $X$ in the process $\Lambda \Lambda \to S X$ is chosen to conserve strong isospin~\footnote{We thank the authors of \cite{KolbTurnerdraft} for pointing out that strong isospin forbids $\Lambda \Lambda \to S \pi$.}. We assume $X=\gamma$ in what follows, and discuss the rate in detail \Sec{S-prod}.

As required, baryon number is conserved in \Eqm{diff-eqN}{diff-eqS} since $\dot{\sf B} \propto \dot n_N + \dot n_\Lambda + 2 \dot n_S =0$. We use initial conditions $n_N(t=0)=n_0$, $n_\pi(t=0)=T^3\exp(-m_\pi/T)$, and $n_\Lambda (t=0)= n_S (t=0)=0$. We assume that the core has a constant temperature $T=30\mev$ and is at the nuclear saturation density $n_0=0.16{\rm\,fm}^{-3}$.
The $N\to \Lambda$ and $\Lambda\to N$ transition rates in \Eqs{diff-eqN}{diff-eqL} each contain two contributions. Because the $\pi$ population is Boltzmann suppressed, however, $N \pi \to \Lambda$ is unlikely to be important in this environment. Similarly, one may assume that the $\Lambda \to N$ transition rate $\Gamma_{\Lambda\to N}$ is dominated by the $\Lambda$ lifetime in the medium, denoted $\tilde \tau_\Lambda$. This is true in vacuum, where $\tau_\Lambda  \simeq 2.6\tenx{-10}\s$, but in a dense medium we expect that direct $\Lambda$ decay is affected by Pauli blocking%, since $|\vec p_N| \lesssim k_F$, 
; we find that the decay width is reduced, $\tilde \tau_\Lambda \simeq 4\tau_\Lambda$. Because $N \Lambda$ collisions are so frequent, $\Lambda$ disappearance can be dominated by a process analogous to collisional de-excitation, {\it e.g.}~$N \Lambda \to NN$ may be more rapid than spontaneous decay. For the nucleon densities we consider, $ n_N \langle \sigma_{N \Lambda \to N N} v \rangle \gtrsim \tilde \tau_\Lambda^{-1}$ if $\langle \sigma_{N \Lambda \to N N} v \rangle \gtrsim 10^{-29} \cm^3/\s$.

One important feature of \Eq{diff-eqS} is that $S$ disappearance has only one channel, which is suppressed by the large binding energy of the $S$, since $n_\gamma(E_\gamma > B_S) \sim T^3 \exp(-B_S/T)$ is small. Thus, the same features that guarantee the $S$ is cosmologically metastable ensure that it cannot be efficiently destroyed in the proto-neutron star environment: $S$ decay is doubly weak, and $S$ fission is suppressed by its large binding energy, $B_S \gg T$. For this reason, $S$ acts as a sink for baryon number until $n_S \simeq n_N$. If $S$ formation is efficient, all baryon number in the hot proto-neutron star core will be processed into $S$ particles.

%\begin{figure}[bt]
%\begin{center}
%\includegraphics[width=0.46\textwidth]{sszoom}
%\caption{Contours of $t_S$ as defined in \Eq{S-equilibration}. Above the solid (dashed) line, $S$ production equilibrates in less than 10 s (1 ms).}
%\label{tS-contour}
%\end{center}
%\end{figure}

The $S$ abundance from \Eqm{diff-eqN}{diff-eqS} approximately yields to analytic solution. First, consider the limiting scenario $\langle \sigma_{ \Lambda \Lambda \to S \gamma} v \rangle \to 0$. It is clear that $n_N, n_\Lambda$ reach an equilibrium where $\dot n_\Lambda =\dot n_N = 0$ when the $\Lambda$ abundance has increased to
\beq
\bar n_\Lambda = n_N \frac{  \langle \sigma_{N N \to \Lambda N} v \rangle}{  \langle \sigma_{N \Lambda \to N N} v \rangle+ \tilde \tau_\Lambda^{-1}/n_N}.
\eeq
The $N-\Lambda$ cross sections are related by detailed balance, such that $\bar n_\Lambda / n_N \leq \langle \sigma_{N N \to \Lambda N} v \rangle/\langle \sigma_{N \Lambda \to N N} v \rangle = \pL m_\Lambda/m_N \pR^{3/2} \exp\bL-(m_\Lambda-m_N)/T\bR$.
Next, we note that for constant $n_N$, \Eq{diff-eqL} has an analytic solution even with $\langle \sigma_{ \Lambda \Lambda \to S \gamma} v \rangle \neq 0$:
\alg{ \label{XL0}
n_\Lambda(t) &= \bar n_\Lambda \frac{2  \tanh (\gamma t /2)}{\tanh (\gamma t /2)+ \sqrt{1 + r}},
\\ {\rm with~~}\gamma &\equiv (\tilde \tau_\Lambda^{-1}+n_N  \langle \sigma_{N \Lambda \to N N} v \rangle) \sqrt{1 + r} 
\\ {\rm and~~} r &\equiv \frac{8 \bar n_\Lambda \langle \sigma_{ \Lambda \Lambda \to S \gamma} v \rangle}{\tilde \tau_\Lambda^{-1}+ n_N \langle \sigma_{N \Lambda \to N N} v \rangle}
~.}
The asymptotic $\Lambda$ abundance is $n_\Lambda^\infty \equiv n_\Lambda(t \gg \gamma^{-1}) = 2\bar n_\Lambda/(1+\sqrt{1+r})$, where the time constant satisfies $\gamma^{-1} \leq \tilde \tau_\Lambda$. Crucially for our purposes, this happens promptly on the timescales of relevance for a supernova explosion.

Given $n_\Lambda^\infty$, \Eq{diff-eqS} dictates that the $S$ abundance will rise linearly as long as fission is unimportant, $n_S(t) \simeq \bar n_S(t)\equiv (n_\Lambda^\infty)^2 \langle \sigma_{ \Lambda \Lambda \to S \gamma} v \rangle  t$. This is true until an $\cO(1)$ fraction of baryons are in $S$ dibaryons, which happens at a time $t_S$ defined by $2 n_S(t_S) = n_N(t_S)$%, after which fission from pion collisions becomes important and $n_S$ equilibrates
. We find that $t_S$ defined in this way is equivalent to solving for $\bar n_S(t_S) = n_0$, to an accuracy of 10\%, or
\beq \label{S-equilibration}
t_S = \frac{n_0}{(n_\Lambda^\infty)^2 \langle \sigma_{ \Lambda \Lambda \to S \gamma} v \rangle} .
\eeq
Plugging $n_\Lambda^\infty$ into \Eq{S-equilibration} and assuming a hierarchy of rates: $ \bar n_\Lambda \langle \sigma_{ \Lambda \Lambda \to S \gamma} v \rangle \ll  n_0 \langle \sigma_{N \Lambda \to N N} v \rangle \sim \tilde \tau_\Lambda^{-1}$, we find that $S$ production equilibrates at a time ${t_S \simeq \s \frac{4\tenx{-34}\cm^3/\s}{\langle \sigma_{ \Lambda \Lambda \to S \gamma} v \rangle} \bL 1 + \frac{2\tenx{-32}\cm^3/\s}{ \langle \sigma_{N N \to \Lambda N} v \rangle} \bR^2}$. 
After $t_S$ has elapsed, backreaction will become non-negligible due to the heat dumped by the exothermic $S$ fusion process. Due to the large binding energy, $\gamma S \to \Lambda \Lambda$ will become important only deep in the back-reacted regime. By this time, however, the assumption of thermal equilibrium will have long since broken down, and the proto-neutron star will either combust or decay entirely to $S$ particles.

%The former is a function of inherent properties of the dibaryon; we treat it in detail in the next section. To determine the latter
%We treat these in turn.

\section{$\Lambda$ and $S$ Production}
If $t_S$ given in \Eq{S-equilibration} is short compared to the neutrino burst from SN1987A, which was observed to last for $t_\nu \sim \cO(10\s)$, $S$ production equilibrates quickly on the timescales of relevance to the proto-neutron star. %In \Fig{tS-contour}, we show contours of $t_S$.
As we discuss in the next section, a proto-neutron star composed entirely of $S$ dibaryons is incompatible with observations. Our analysis indicates that for $\langle \sigma_{ \Lambda \Lambda \to S \gamma} v \rangle \gtrsim 10^{-34}\cm^3/\s$, $S$ production is fatal for the proto-neutron star.
Here, we calculate $\langle \sigma_{N N \to \Lambda N} v \rangle$ and $\langle \sigma_{ \Lambda \Lambda \to S \gamma} v \rangle$.

\subsection{$\Lambda$ Production Cross Section}

To obtain $\langle \sigma_{N N \to \Lambda N} v \rangle$, we first observe that all rates $N \dots \leftrightarrow \Lambda \cdots$ share a strangeness-changing coupling $g_{\Lambda N \pi}$. We obtain this coupling from the in-vacuum $\Lambda$ lifetime, 
\beq \label{tau-life}
\tau_\Lambda^{-1} \simeq \Gamma_{\Lambda \to N \pi} \simeq \frac{g_{\Lambda N \pi}^2}{8\pi m_\Lambda} \frac{|\vec p_N|}{m_\Lambda} \bL \pL m_\Lambda -m_N \pR^2 -m_\pi^2 \bR,
\eeq
giving $g_{\Lambda N \pi}^2 \simeq 7\tenx{-11}$. Because strangeness-changing processes are weak processes, this small dimensionless number can be interpreted as coming from $(G_F m_N^2)^2 \sim 10^{-10}$.
Assuming a constant matrix element, appropriate in the limit of small $m_\pi$ \cite{Brinkmann:1988vi, Raffelt:1993ix, Raffelt:2006cw}, and assuming that the momentum released to the nucleons is large compared to the Fermi momentum, we may write $\langle \sigma_{N N \to \Lambda N} v \rangle \equiv a g_{\Lambda N \pi}^2 \alpha_{N \pi} \sqrt{T/\pi m_N^3 m_\Lambda^2} \simeq  a \tenx{-27} \cm^3/\s,$ where $\alpha_{N \pi}\simeq 15$ and $a$ is a function of temperature and density that parameterizes our ignorance of complicated, higher-order physics that may become important in the proto-neutron star environment%, but which in principle may be calculated in chiral perturbation theory
. A more complete calculation including the effects of nucleon degeneracy, described in \App{nncalc}, gives $a \simeq 0.3-0.5$ for the temperatures and densities of interest if single-pion exchange is a good description of the scattering.

It is well known that pion exchange is nonperturbative, so it is possible that higher-order diagrams have a non-negligible interference with the tree-level scattering. If there is a cancellation to $10\%$ in the matrix element, then $a \simeq 10^{-2}$, and the cross section is $\langle \sigma_{N N \to \Lambda N} v \rangle \simeq 10^{-29}\cm^3/\s$. To be conservative, we will use $\langle \sigma_{N N \to \Lambda N} v \rangle = 3\tenx{-30}\cm^3/\s$ as a default value for the rest of this note, corresponding to a 10\% cancellation in the matrix element for this process that is sustained for the entirety of the proto-neutron star explosion, on top of the $\sim \cO(50\%)$ suppression from $m_\pi$-effects and nucleon degeneracy. We emphasize that, although such cancellations are known to exist at the $\sim \cO(50\%)$ level in the context of $N-N$ scattering, a cancellation of $\sim \cO(90\%)$ would be extremely unusual. But a larger value of $\langle \sigma_{N N \to \Lambda N} v \rangle$ will hasten the rate at which baryon number is processed into $S$ particles, so we choose this value to ensure that our results are indeed conservative.

We also mention here that we have neglected additional baryon species. This is reasonable because baryons of increasing strangeness are increasingly massive. For instance, the equilibrium $\Xi$ population experiences a Boltzmann suppression such that $n_\Xi n_N \lesssim (n_\Lambda^\infty)^2$. %Thus, direct $S$ production via $N \Xi \to S \pi$ is somewhat subdominant to the processes we calculate, but because $\Xi$ and $N$ are not strong isospin singlets, this purely strong process may be enhanced compared to the electroweak channels we consider.
 Including such additional baryons would marginally increase the $S$ production rate, but more importantly would make the cancellation we implicitly absorb even more unlikely. Thus, our analysis is conservative, but this contributes subdominantly to the calculation of $t_S$.

%
%\begin{figure*}[t]
%\begin{center}
%\includegraphics[height=0.27\textheight]{melsqscaled} ~~~~~~~~
%\includegraphics[height=0.27\textheight]{melsq}
%\caption{{\bf Left:} $g_{\Lambda S}^2$ scaled by an overall geometric factor, as a function of the $\Lambda$ charge radius. {\bf Right:} $g_{\Lambda S}^2$ as a function of $r_S$ for two different values of the $\Lambda$ charge radius.}
%\label{fraction_NN}
%\end{center}
%\end{figure*}

\subsection{$S$ Production Cross Section}
\label{S-prod}

We now calculate the cross section for $\Lambda \Lambda \to S \gamma$. Given the range of dibaryon masses considered, this process is exothermic and involves no change of strangeness. The effective Lagrangian that allows this process is
\beq \label{S-lag}
\cL \supset d_\Lambda \bar \Lambda \sigma^{\mu \nu} \Lambda F_{\mu \nu} + g_{\Lambda S} \oL{\Lambda^c} \Lambda S^\dagger +{\rm h.c.},
\eeq
where the dipole moment $d_\Lambda =-0.613 \pm 0.001 \mu_N \simeq (10^4 \mev)^{-1}$, $\Lambda^c$ is the $\Lambda$ charge conjugate, and $g_{\Lambda S}$ is a function of inherent dibaryon properties discussed in more detail below. From direct calculation, we find that for the temperatures and binding energies of interest the cross section due to the Lagrangian in \Eq{S-lag} is
\alg{
\langle \sigma_{\Lambda \Lambda \to S \gamma} v \rangle &\simeq  3\tenx{-23} \frac{g_{\Lambda S}^2 B_S}{176.9 \mev} \frac T{30 \mev} \frac{\cm^3}\s,
}
where we have assumed that the fraction of final states with the quantum numbers of the $S$ is $1/1440$. The magnitude of $g_{\Lambda S}$ introduces the largest uncertainty into our calculations.

The coupling $g_{\Lambda S}$ is in principle a low-energy output of QCD. Since strongly coupled QCD is not currently amenable to analytic calculation, and since lattice studies are difficult for a large number of light quarks, we must choose a model to calculate $g_{\Lambda S}$. In prior work, $g_{\Lambda S}$ has been determined by a geometric factor given by the integrated wavefunction overlap~\cite{Farrar:2003qy, Gross:2018ivp}. We will follow these works and use the Isgur-Karl~\cite{Isgur:1978wd} and Brueckner-Bethe-Goldstone~\cite{DAY:1967aa} models to calculate the overlap of the $\Lambda$s and the $S$. This is, of course, only one model of the complicated nuclear quantum mechanics involved.%, and this coupling can in principle be calculated a bag model if so desired.

As discussed in more detail in \App{wfcalc}, the wavefunction overlap has a striking dependence on the dibaryon radius, $r_S$, and the $\Lambda$ radius, $r_\Lambda$. The $S$ radius is entirely unknown, so to be maximally conservative we simply require that $r_S$ exceed the Compton wavelength of the dibaryon plus some fraction $x$ of the Compton wavelength of the lightest meson to which it couples, as advocated in \cite{Farrar:2018hac}. This gives
\beq \label{rsmin}
r_S \geq \frac1{m_S} + \frac{x}{m_{f^0}} = 0.1 \fm \frac{2054 \mev}{m_S} + 0.34x\fm%\bL 2\Lambda_{\rm QCD} m_S\bR^{-1/2} = 0.22\fm \sqrt{2054\mev/m_S}
.
\eeq
We will show results for $x=0,0.1$ in our final plots. Since the dibaryon is a boson, it has no inherent exclusion principle to provide pressure against collapse, so a large coupling to a vector mediator satisfying $g_\omega /m_\omega \geq g_\sigma/m_\sigma$ is necessary \cite{Faessler:1997jg}. We return to this point below.
If instead we required that the non-relativistic zero-point kinetic energy, $r^{-2}/2m$, of quarks localized within the dibaryon of radius $r_S$ should not exceed the energy scale of QCD confinement, we would find a sharper bound. Asserting only that $m_q \leq m_S$ would translate to a bound $r_S \geq 0.22\fm \sqrt{2054\mev/m_S}$. Taking a constituent quark mass $m_q \simeq m_S/6 $, we would have $r_S \gtrsim 0.53\fm$. This latter value roughly matches the constituent quark Compton wavelength, $6/m_S \gtrsim 0.58\fm$. For this reason, restricting to the range $0.1 \fm \leq r_S \leq 1.0\fm$ is very conservative, and the choice $r_S \simeq 0.1\fm$ would be an extremely novel feature for a QCD bound state.

Likewise, the $\Lambda$ radius carries some uncertainty. It is reasonable to assume that increasing strangeness leads to a more compact baryon, $r_\Lambda \lesssim r_N$. The strong interaction radius extracted from experimental data $\sqrt{\vev{r_\Lambda^2}_{\rm st}} = 0.76\pm0.01 \fm$~\cite{Povh:1990ad} is somewhat larger than the na\"ive value in the constituent quark model, $r_\Lambda \simeq \bL 2 \Lambda_{\rm QCD} m_\Lambda/3 \bR^{-1/2} \simeq 0.51\fm$. %The wavefunction overlap is maximized when $r_\Lambda \simeq r_S$, and the greatest suppression occurs if $r_\Lambda \gg r_S$.
Being cautious once again, we decide to show the relatively wide range $0.5 \leq r_\Lambda \leq 0.8\fm$, where the lower limit is chosen to account for the possibility that the $\Lambda$ charge radius is smaller than the strong interaction radius.

Finally, we note that if the binding energy is near the extreme of the range in \Eq{bs-range}, then the process $\Lambda \Lambda \to S \pi \pi$ is on-shell and exothermic as well. Emission of two pions is likely dominated by quark rearrangement processes, which occur at long distances due to the small pion mass. Because the light quarks in the initial state can escape to distances of order the pion Compton wavelength, the cross section should be $\langle \sigma_{\Lambda \Lambda \to S \pi \pi} v \rangle \sim \cO( m_\pi^{-2}),$ which does not suffer from an exponential wavefunction overlap suppression factor. There will be $\sim \cO(0.1)$ hadronization and mass-dependent phase-space suppression factors that we cannot calculate, however. Regardless, for masses $m_S \lesssim 1950 \mev$, we expect that the timescale $t_S \ll \text{ns}$ is unsuppressed and independent of $r_S$. This strengthens the argument considerably in the mass range $1850-1900 \mev$, which is of particular interest in recent studies \cite{Farrar:2018hac}.

\begin{figure*}[t]
\begin{center}
\includegraphics[width=0.48\textwidth]{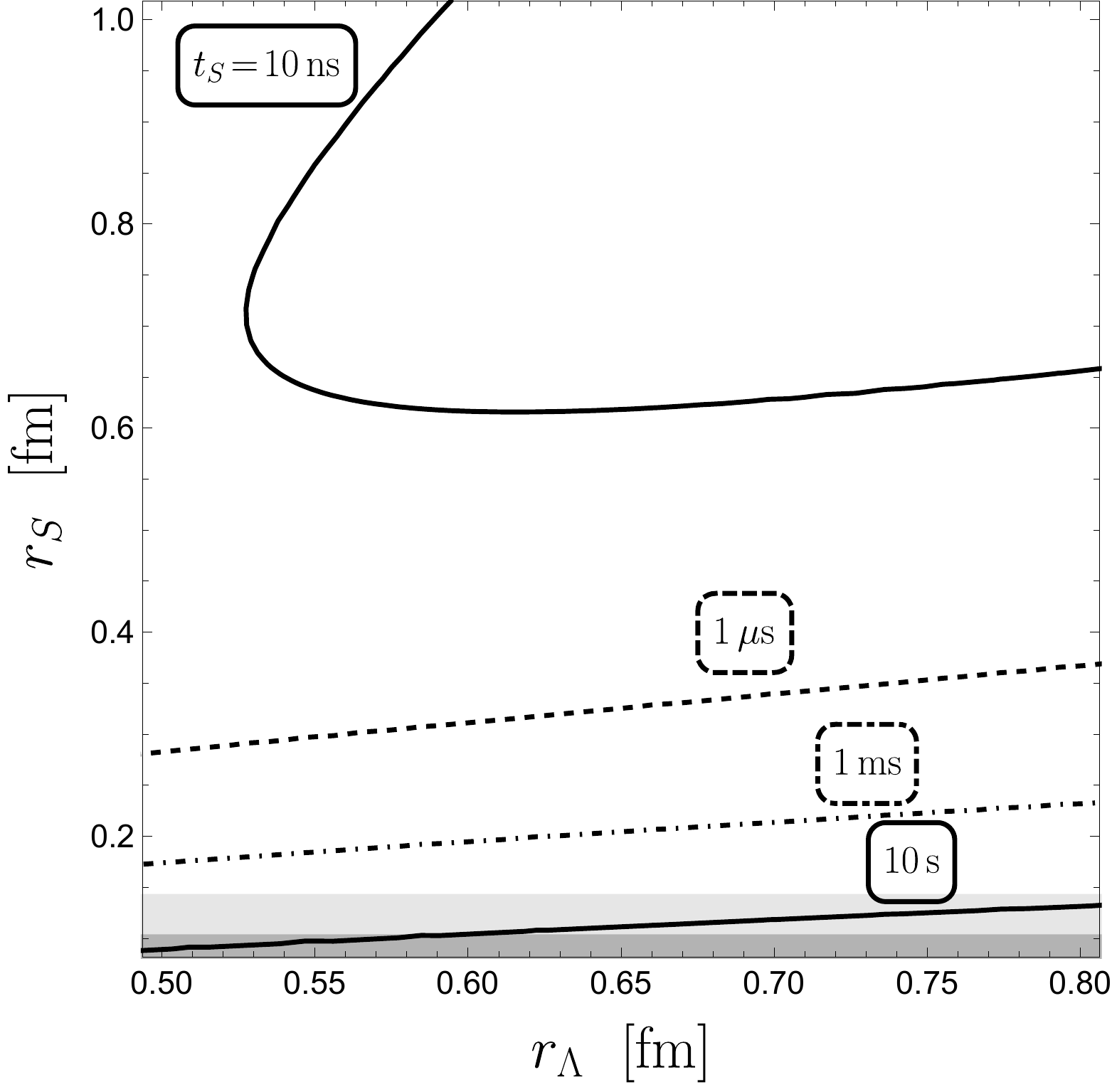}~~~~
\includegraphics[width=0.48\textwidth]{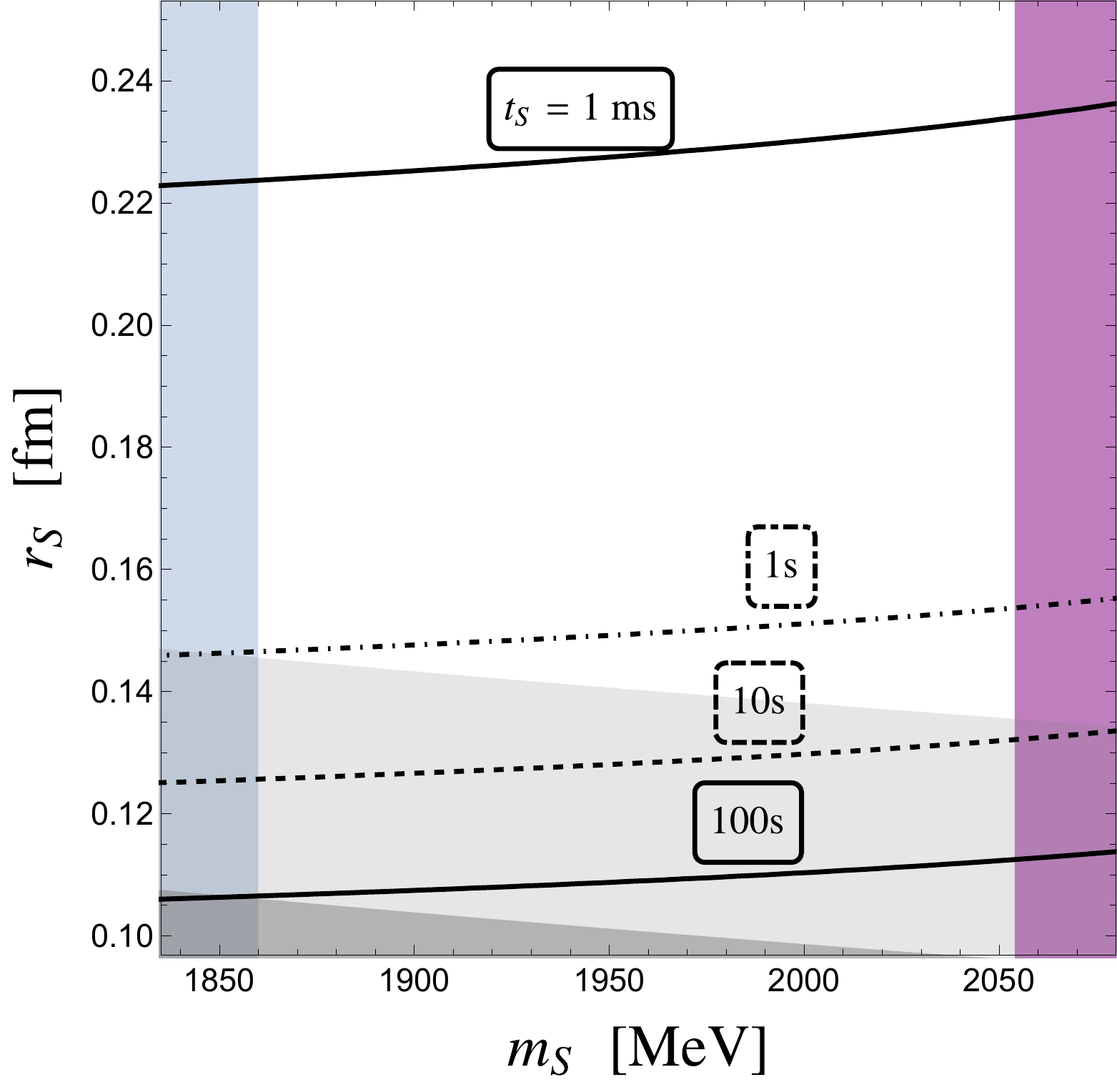}
\caption{{\bf Left:} Contours of $t_S$ as defined in \Eq{S-equilibration} for $m_S=1900\mev$ as a function of the $\Lambda$ and $S$ sizes. The gray region violates \Eq{rsmin} for $x=0,0.1$. {\bf Right:} Contours of $t_S$ for $r_\Lambda=0.76\fm$. In both panels, we have assumed $\langle \sigma_{N \Lambda \to N N} v \rangle = 3\tenx{-30} \cm^3/\s$. The gray region violates \Eq{rsmin} for $x=0,0.1$. In the blue region, ${}^{16}$O nuclei are destabilized. In the purple region, the dibaryon has a singly weak decay. All of the parameter space depicted in each panel has $t_S \ll 10 \s$, and is thus ruled out by the observation that SN1987A continued to emit neutrinos for $t_\nu \simeq 10\s$, unless $r_S$ is very close the minimum value allowed by \Eq{rsmin}.}
\label{mf-plane}
\end{center}
\end{figure*}

\section{Fate of the Proto-Neutron Star}

We show our final results in \Fig{mf-plane}, fixing $\langle \sigma_{N \Lambda \to N N} v \rangle = 3\tenx{-30} \cm^3/\s$. %Without any cancellation in this matrix element, our results would be strengthened by orders of magnitude.
The left panel of \Fig{mf-plane} depicts the lifetime as a function of $r_\Lambda$ and $r_S$ for $m_S=1900\mev$. %As discussed in the preceding section, these radii are expected on theoretical grounds to be near $0.6\fm$ and $0.76\fm$. 
In the dark (light) gray region, $r_S$ violates \Eq{rsmin} for $x=0 (0.1)$. $S$ production equilibrates in the proto-neutron star much faster than $10\s$ for most of the range of $r_\Lambda$ and $r_S$ that we consider, unless $r_S$ is very close to $0.1\fm$. For such a small radius, the coupling can be as small as $g_{\Lambda S}^2 \sim 10^{-11} - 10^{-14}$ by the wavefunction overlap calculation discussed in \App{wfcalc}.

In the right panel of \Fig{mf-plane}, we depict $t_S$ for $r_\Lambda=0.76\fm$ as a function of dibaryon mass $m_S$ and radius $r_S$. In the blue shaded region, and at smaller masses, the existence of an $S$ dibaryon renders ${}^{16}$O nuclei unstable~\cite{Gross:2018ivp}. In the purple shaded region, and at larger masses, the dibaryon cannot possibly be cosmologically metastable, since it has a singly weak decay~\cite{Farrar:2003qy}. In the dark (light) gray region, $r_S$ violates \Eq{rsmin} for $x=0 (0.1)$. 

In all of the heretofore phenomenologically viable parameter space, we find that $t_S \ll 10\s$, unless $r_S$ is very close to $0.1\fm$.
Such a fast equilibration of the $S$ number density implies that all baryons in the proto-neutron star interior rapidly find themselves inside $S$ dibaryons. This would have catastrophic consequences. Since the $S$ dibaryon is a compact boson, its equation of state would be characterized by a pressure that is much smaller than the pressure of the neutron-rich matter it replaces. Fermi degeneracy and strong interactions between neutrons produce enough pressure to support neutron stars up to a maximum mass $> 2$ $M_\odot$, compatible with observations of massive neutron stars~\cite{Demorest:2010bx, Antoniadis:2013pzd}. In contrast, matter composed of the $S$ dibaryon, where pressure is solely due to short-range repulsion, would be too compressible to support such a large maximum mass.  We have estimated the strength of repulsive interactions needed to support a maximum mass of $2~M_\odot$ and found that, in a simple model where dibaryons interact by exchanging vector mesons with mass $m \simeq m_\omega ~ 800$ MeV, the coupling strength needed to produce adequate repulsion to support observed neutron star masses is unnaturally large. Treating the dimensionless dibaryon-vector meson coupling strength $g_S$ as free parameter we calculated the equation of state of the interacting dibaryon system in mean field theory and found that to support a maximum mass $> 2~M_\odot$ we require unnaturally large  values of $g_S>10$. As discussed above, a coupling large enough to ensure stability would also increase the characteristic size of the dibaryon, and would preclude $r_S \simeq 1/m_S$. Interestingly, in this simple model with large repulsive couplings we also find that the radius of typical neutron stars (with masses in the range $1.2-1.5 M_\odot$) would be greater than 15 km. This is conflict with the constraints from GW1701817 \cite{Bauswein:2017vtn, Annala:2017llu, Most:2018hfd}. Taken together, this suggests that interactions between dibaryons is unlikely to change our conclusion that the star composed mostly of tightly bound dibaryons is incompatible with observations. 

Finally, the large energy released by the exothermic reactions, $B_S \sim 100\mev$ per baryon, is comparable to the gravitational binding energy. $S$ production likely unbinds the stellar remnant, but even if the proto-neutron star remains intact, this heat dump disrupts the standard evolution of the proto-neutron star. 

%For instance, the fact that $S$ dibaryons do not couple to a weak current \sr{Is this correct? The weak neutral current couples to the flavor singlet current} is certainly in conflict with the fact that a neutrino cooling phase of $\sim \cO(10\s)$ was observed from the hot interior of Supernova 1987A \cite{Hirata:1987hu, Bionta:1987qt, Alekseev:1987ej}. 

%In conclusion, we robustly find that all baryon number is processed into $S$ dibaryons during the proto-neutron star explosion for any reasonable values of the baryon and dibaryon cross sections. Since the values in our analysis are quite conservatively chosen to be well within the experimentally and theoretically viable ranges for these parameters, we conclude that the observation of a core collapse supernova that cooled over the span of several seconds rules out the existence of the $S$ dibaryon.

\section{Conclusions}

In this work, we have shown that the hot interior of a proto-neutron star provides a valuable laboratory for probing the nature of the proposed deeply bound $S$ dibaryon. The $S$ can be produced on shell in $\Lambda \Lambda$ collisions, and this exothermic reaction equilibrates quickly on the timescales of relevance to the neutron star explosion unless the dibaryon production cross section is suppressed by 11 orders of magnitude. In the context of a wavefunction overlap calculation, we find that this is possible only if the $S$ radius is very close to its Compton wavelength $\simeq 0.1\fm$. Absent this suppression, rapid equilibration of $S$ density implies that all baryon number inside of the proto-neutron star is processed into $S$ number much more quickly than the observed neutrino burst of Supernova 1987A. Indeed, the energy released in the hard gamma rays that accompany the formation of an $S$ is large could unbind the proto-neutron star entirely. Finally, if such an object were to survive, an entire star composed entirely of $S$ particles would have a much softer equation of state than a neutron star. Thus, the existence of proto-neutron stars and old neutron stars with properties roughly similar to those predicted from standard nuclear astrophysics seems to be in grave tension with the presence of a dibaryon in the QCD spectrum.

\acknowledgements
We thank Nikita Blinov, Glennys Farrar, Rocky Kolb, and Michael Turner for discussions.  SDM was supported by Fermi Research Alliance, LLC under Contract No.~DE-AC02-07CH11359 with the U.S. Department of Energy, Office of Science, Office of High Energy Physics. The United States Government retains and the publisher, by accepting the article for publication, acknowledges that the United States Government retains a non-exclusive, paid-up, irrevocable, world-wide license to publish or reproduce the published form of this manuscript, or allow others to do so, for United States Government purposes. SR and SS are supported by Department of Energy grant DE-FG02-00ER41132.

~\\ \noindent {\it Note Added:} As our paper was being finalized, we received a draft of \cite{KolbTurnerdraft}, which critically addresses the possibility that the $S$ can attain an interesting cosmological abundance. The underlying assumption of \cite{KolbTurnerdraft} is that $S$ is present in the QCD spectrum, which makes it complementary to the present work. While this work was in review, we also became aware of \cite{Echenard:2018rfe}, which finds no candidate events from a search for the $S$ in $\Upsilon$ decays.

\appendix

\begin{figure*}[t]
\begin{center}
\includegraphics[height=0.285\textheight]{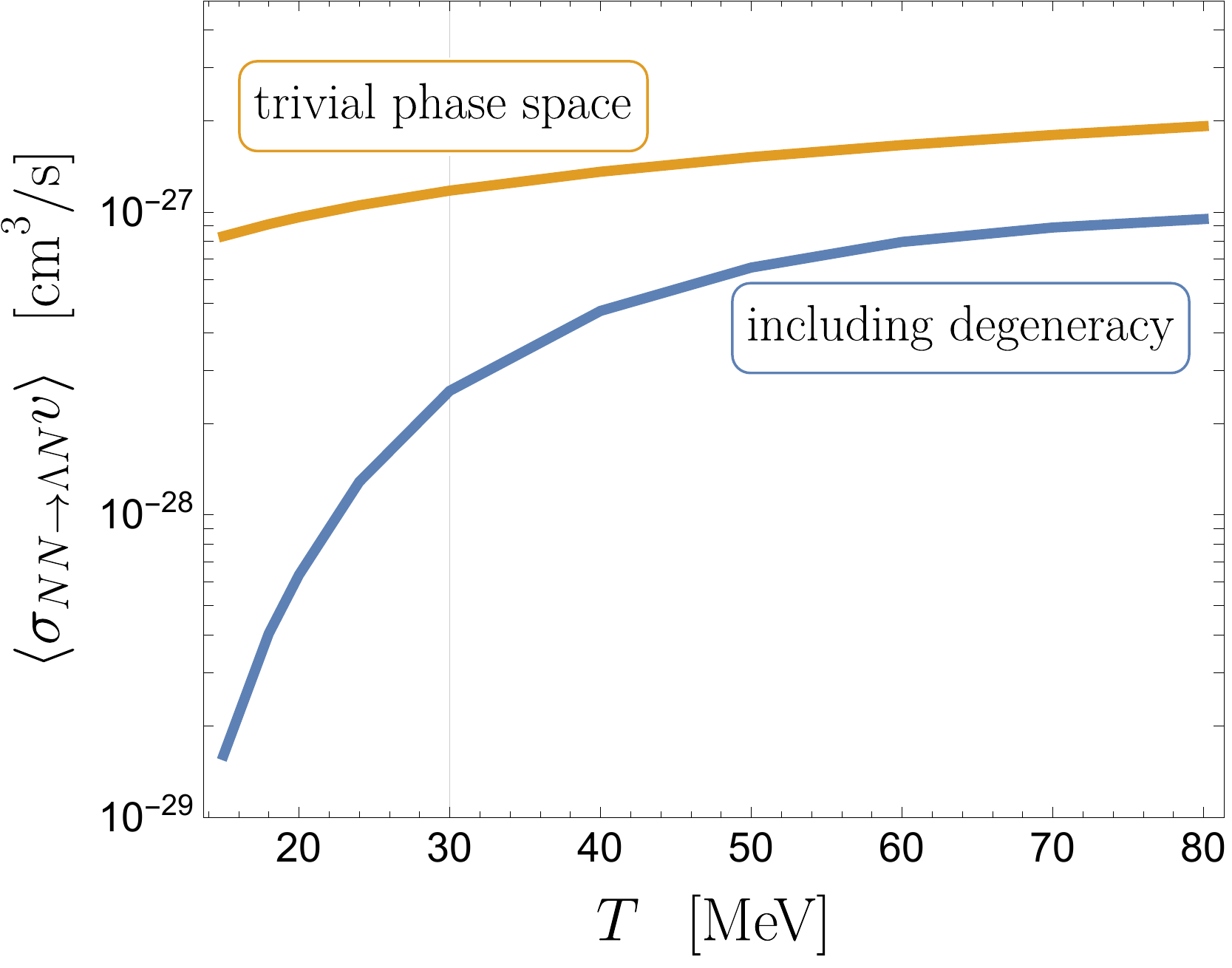}~~~~~~
\includegraphics[height=0.285\textheight]{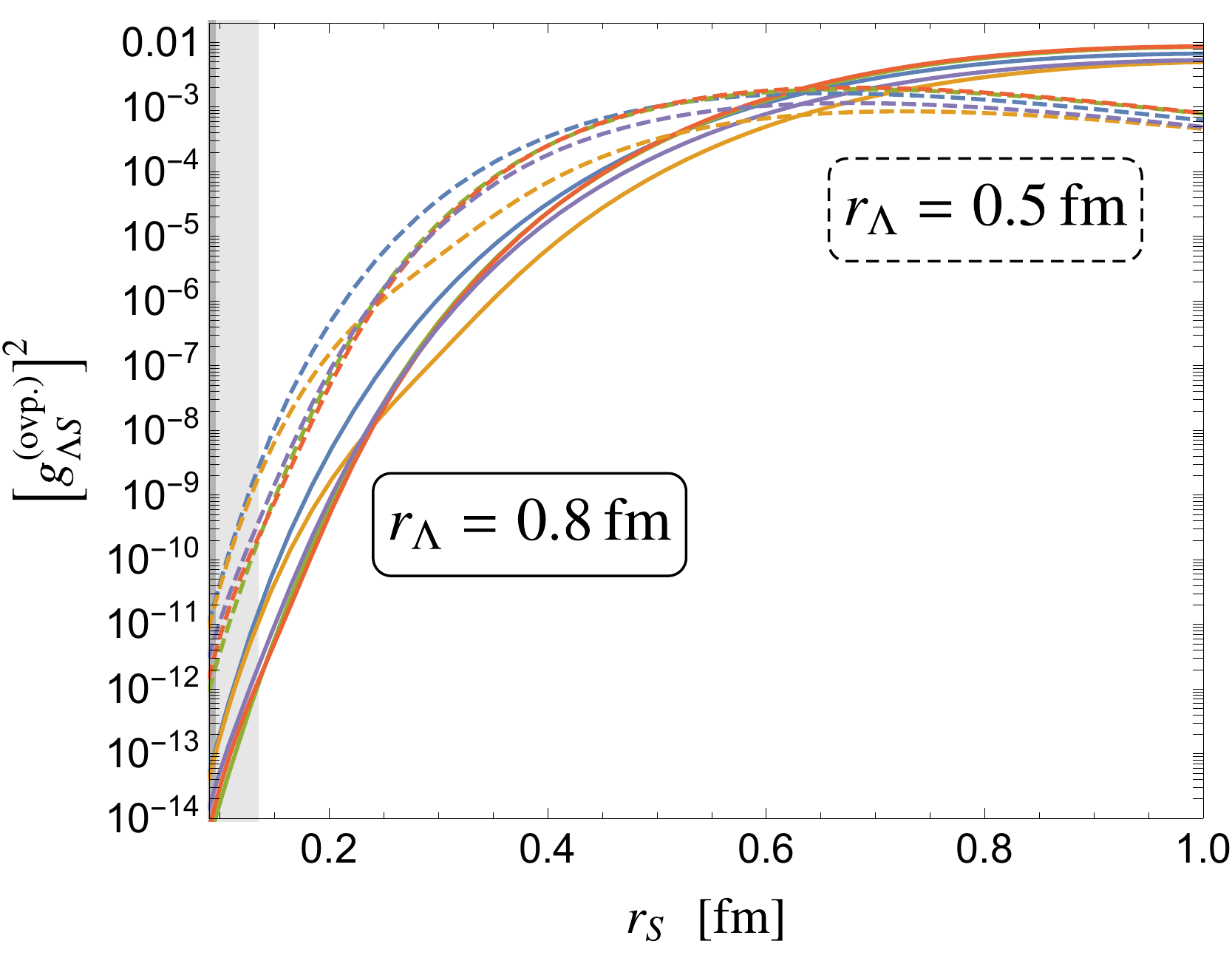}
\caption{{\bf Left:} The value of $\langle \sigma_{NN \to \Lambda N} v\rangle$ with and without phase space degeneracy effects. {\bf Right:} The coupling $g_{\Lambda S}^2$ from wavefunction overlap, as a function of $r_S$ for two different values of $r_\Lambda$, as in \Eq{wfovp}. In the gray region, $r_S$ violates \Eq{rsmin} for $x=0.1$.}
\label{melsq}
\end{center}
\end{figure*}

\section{$NN \to \Lambda N$ Calculation} \label{nncalc}
The cross section for $\langle \sigma_{NN \to \Lambda N} v\rangle$ determines the equilibrium $\Lambda$ abundance, which in turn determines $t_S$. Assuming a trivial matrix element for single pion exchange and integrating over non-degenerate phase space, in agreement with calculations of nucleon-nucleon scattering in the single-pion-exchange limit \cite{Brinkmann:1988vi, Raffelt:1993ix, Raffelt:2006cw}, gives $\langle \sigma_{N \Lambda\to N N} v \rangle \equiv g_{\Lambda N \pi}^2 \alpha_{N \pi} \sqrt{T/\pi m_N^3 m_\Lambda^2} \simeq \tenx{-27} \cm^3/\s,$ where $g_{\Lambda N \pi}$ is obtained from \Eq{tau-life} and $\alpha_{N \pi}\simeq 15$. Effects of degeneracy are expected to be mild in this environment \cite{Brinkmann:1988vi}, but should have effects at the $\sim \cO(1)$ level \cite{Raffelt:1993ix}. Here we confirm this expectation with explicit calculation.

The rate per unit volume for production of $\Lambda$ baryons in $NN$ collisions is
\alg{
\frac\Gamma{\rm Vol} = \int \prod_{i=1}^4 \frac{d^3 \vec p_i}{(2\pi)^3 2E_i}  (2\pi)^4 \delta^{(4)}(p_1+p_2 - p_3 -p_4)  \times
\\ \times f(N_1) f(N_2) \bL 1-f(N_3 )\bR  \bL 1-f(\Lambda_4) \bR \mL \cM_{N N \to \Lambda N} \mR^2,
}
where $f(B_i) = \cbL \exp\bL(E_i - \mu_i)/T\bR +1 \cbR^{-1}$ is the Fermi-Dirac distribution function for the baryon $i$. The matrix element $\cM_{N N \to \Lambda N}$ follows from the Lagrangian $\cL \supset g_{NN\pi} \bar N \gamma_5 N \pi + g_{\Lambda N\pi} \bar \Lambda \gamma_5 N \pi +{\rm(h.c.)}$, where $g_{NN\pi}$ is given by the Goldberger-Treiman relation. The chemical potential and temperature are related by the requirement that $n_0 = \int \frac{2 \times d^3 \vec p}{(2\pi)^3} f(N_i)$%, where the factor of 2 comes from the spin degeneracy
. The chemical potentials satisfy $\mu_\Lambda = \mu_N$ by detailed balance. We find that $\mu_N \gtrsim m_N$, and thus the $N$ are mildly degenerate, for $T \lesssim 50\mev$.

Because the nucleon densities are fixed to the saturation value, we may determine the cross section by
\beq \label{NN-exact}
\langle \sigma_{NN \to \Lambda N} v\rangle = \frac\Gamma{\rm Vol} n_0^{-2}.
\eeq
We plot the results of \Eq{NN-exact} and the value $g_{\Lambda N \pi}^2 \alpha_{N \pi} \sqrt{T/\pi m_N^3 m_\Lambda^2} \simeq \sqrt{T/30\mev}\tenx{-27} \cm^3/\s$ for $15\mev \leq T \leq 80 \mev$ in \Fig{melsq}, left panel. The result with the assumption of a trivial phase space is a factor of $\sim 3$ higher at $T=30\mev$. The discrepancy shrinks at large $T$, where corrections due to $m_\pi \neq0$ are less important.

\section{Wavefunction Overlap Calculation} \label{wfcalc}

Following \cite{Farrar:2003qy}, we integrate the Isgur-Karl wavefunctions of two initial-state baryons against a relative wavefunction that incorporates the $\Lambda-\Lambda$ potential. In agreement with \cite{Farrar:2003qy, Gross:2018ivp}, we have
\alg{ \label{wfovp}
g_{\Lambda S}^{\rm (ovp.)} = 32 \pL \frac3{2\pi} \pR^{3/4} \frac{(r_S/r_\Lambda)^{9/2}}{\bL 1+(r_S/r_\Lambda)^2 \bR^6 } r_\Lambda^{-3/2} \times 
\\ \times \int d^3 a \, \psi_{\rm rel} \psi_\gamma \exp^{-3a^2/4r_S^2},
}
where $\psi_{\rm rel}$ has mass dimension $-3/2$. We assume that the $\gamma$ is a plane wave whose presence allows conservation of energy and momentum. It is possible that in processes where strong mesons are emitted, such as $\Lambda \Lambda \to S \pi \pi$ or $N \Xi \to S \pi$, the presence of the $\pi$ has qualitative significance for the process of $S$ formation. For instance, if quark rearrangement is important, then some of the quarks in the initial state may escape to the $\pi$, which is at a distance much larger than $r_S$, meaning that the wavefunctions need not coincide as exactly as in our model calculation, and the cross section may be as large as $m_\pi^{-2}$. However, such effects are difficult to quantify in the absence of a calculable model of hadronization, so we restrict to $\Lambda \Lambda \to S \gamma$, where such considerations are irrelevant. Nonetheless, we stress that a complete picture should include all rearrangement effects, and may lead to substantially larger cross sections.

For numerical values of $\psi_{\rm rel}$, we use the relative wavefunctions depicted in Fig.~5 of \cite{Morita:2014kza}. These wavefunctions are generated from potentials calibrated on the Nagara event, which requires a slightly repulsive interaction. The inverse scattering length is small and negative, while consistency should require that the inverse scattering length for a very deeply bound dibaryon is large and positive %, as may be indicated by recent chiral effective field theory results
\cite{Haidenbauer:2015zqb, Li:2018tbt}. Needless to say, an attractive potential would lead to a relative wavefunction that was larger near the origin. On the other hand, $\Lambda \leftrightarrow N$ transitions can occur more quickly than $\Lambda \Lambda$ fusion for small $g_{\Lambda S}$, meaning that the two baryons involved in a single $\Lambda \Lambda \to S \gamma$ event may change strangeness while they are within range of each other's potential. Thus, the correct relative wavefunction may be a linear combination of relative $\Lambda-N$ and $\Lambda-\Lambda$ wavefunctions. For this reason, the slightly repulsive potentials of \cite{Morita:2014kza} provide a conservative model of this process.

We show the final results of integrating \Eq{wfovp} in \Fig{melsq}. As is clear, $g_{\Lambda S}$ calculated in this way is largely insensitive to the details of the wavefunctions: all of these relative wavefunctions integrate to $\cO(1)$ numbers. The more important scaling has to do with the large polynomial dependence on $r_S$ and $r_\Lambda$ and the exponential dependence on $r_S$, which cause the square of the overlap to vary by approximately three orders of magnitude.

\bibliography{eos-b}

%merlin.mbs apsrev4-1.bst 2010-07-25 4.21a (PWD, AO, DPC) hacked
%Control: key (0)
%Control: author (8) initials jnrlst
%Control: editor formatted (1) identically to author
%Control: production of article title (-1) disabled
%Control: page (0) single
%Control: year (1) truncated
%Control: production of eprint (0) enabled
\begin{thebibliography}{43}%
\makeatletter
\providecommand \@ifxundefined [1]{%
 \@ifx{#1\undefined}
}%
\providecommand \@ifnum [1]{%
 \ifnum #1\expandafter \@firstoftwo
 \else \expandafter \@secondoftwo
 \fi
}%
\providecommand \@ifx [1]{%
 \ifx #1\expandafter \@firstoftwo
 \else \expandafter \@secondoftwo
 \fi
}%
\providecommand \natexlab [1]{#1}%
\providecommand \enquote  [1]{``#1''}%
\providecommand \bibnamefont  [1]{#1}%
\providecommand \bibfnamefont [1]{#1}%
\providecommand \citenamefont [1]{#1}%
\providecommand \href@noop [0]{\@secondoftwo}%
\providecommand \href [0]{\begingroup \@sanitize@url \@href}%
\providecommand \@href[1]{\@@startlink{#1}\@@href}%
\providecommand \@@href[1]{\endgroup#1\@@endlink}%
\providecommand \@sanitize@url [0]{\catcode `\\12\catcode `\$12\catcode
  `\&12\catcode `\#12\catcode `\^12\catcode `\_12\catcode `\%12\relax}%
\providecommand \@@startlink[1]{}%
\providecommand \@@endlink[0]{}%
\providecommand \url  [0]{\begingroup\@sanitize@url \@url }%
\providecommand \@url [1]{\endgroup\@href {#1}{\urlprefix }}%
\providecommand \urlprefix  [0]{URL }%
\providecommand \Eprint [0]{\href }%
\providecommand \doibase [0]{http://dx.doi.org/}%
\providecommand \selectlanguage [0]{\@gobble}%
\providecommand \bibinfo  [0]{\@secondoftwo}%
\providecommand \bibfield  [0]{\@secondoftwo}%
\providecommand \translation [1]{[#1]}%
\providecommand \BibitemOpen [0]{}%
\providecommand \bibitemStop [0]{}%
\providecommand \bibitemNoStop [0]{.\EOS\space}%
\providecommand \EOS [0]{\spacefactor3000\relax}%
\providecommand \BibitemShut  [1]{\csname bibitem#1\endcsname}%
\let\auto@bib@innerbib\@empty
%</preamble>
\bibitem [{\citenamefont {Jaffe}(1977)}]{Jaffe:1976yi}%
  \BibitemOpen
  \bibfield  {author} {\bibinfo {author} {\bibfnamefont {R.~L.}\ \bibnamefont
  {Jaffe}},\ }\href {\doibase 10.1103/PhysRevLett.38.195} {\bibfield  {journal}
  {\bibinfo  {journal} {Phys. Rev. Lett.}\ }\textbf {\bibinfo {volume} {38}},\
  \bibinfo {pages} {195} (\bibinfo {year} {1977})},\ \bibinfo {note} {[Erratum:
  Phys. Rev. Lett.38,617(1977)]}\BibitemShut {NoStop}%
%%CITATION = PRLTA,38,195;%%
\bibitem [{\citenamefont {Belz}\ \emph {et~al.}(1996)\citenamefont {Belz} \emph
  {et~al.}}]{Belz:1995nq}%
  \BibitemOpen
  \bibfield  {author} {\bibinfo {author} {\bibfnamefont {J.}~\bibnamefont
  {Belz}} \emph {et~al.} (\bibinfo {collaboration} {BNL-E888}),\ }\href
  {\doibase 10.1103/PhysRevC.56.1164, 10.1103/PhysRevLett.76.3277} {\bibfield
  {journal} {\bibinfo  {journal} {Phys. Rev. Lett.}\ }\textbf {\bibinfo
  {volume} {76}},\ \bibinfo {pages} {3277} (\bibinfo {year} {1996})},\ \bibinfo
  {note} {[Phys. Rev.C56,1164(1997)]},\ \Eprint
  {http://arxiv.org/abs/hep-ex/9603002} {arXiv:hep-ex/9603002 [hep-ex]}
  \BibitemShut {NoStop}%
%%CITATION = HEP-EX/9603002;%%
\bibitem [{\citenamefont {Alavi-Harati}\ \emph {et~al.}(2000)\citenamefont
  {Alavi-Harati} \emph {et~al.}}]{AlaviHarati:1999ds}%
  \BibitemOpen
  \bibfield  {author} {\bibinfo {author} {\bibfnamefont {A.}~\bibnamefont
  {Alavi-Harati}} \emph {et~al.} (\bibinfo {collaboration} {KTeV}),\ }\href
  {\doibase 10.1103/PhysRevLett.84.2593} {\bibfield  {journal} {\bibinfo
  {journal} {Phys. Rev. Lett.}\ }\textbf {\bibinfo {volume} {84}},\ \bibinfo
  {pages} {2593} (\bibinfo {year} {2000})},\ \Eprint
  {http://arxiv.org/abs/hep-ex/9910030} {arXiv:hep-ex/9910030 [hep-ex]}
  \BibitemShut {NoStop}%
%%CITATION = HEP-EX/9910030;%%
\bibitem [{\citenamefont {Kim}\ \emph {et~al.}(2013)\citenamefont {Kim} \emph
  {et~al.}}]{Kim:2013vym}%
  \BibitemOpen
  \bibfield  {author} {\bibinfo {author} {\bibfnamefont {B.~H.}\ \bibnamefont
  {Kim}} \emph {et~al.} (\bibinfo {collaboration} {Belle}),\ }\href {\doibase
  10.1103/PhysRevLett.110.222002} {\bibfield  {journal} {\bibinfo  {journal}
  {Phys. Rev. Lett.}\ }\textbf {\bibinfo {volume} {110}},\ \bibinfo {pages}
  {222002} (\bibinfo {year} {2013})},\ \Eprint {http://arxiv.org/abs/1302.4028}
  {arXiv:1302.4028 [hep-ex]} \BibitemShut {NoStop}%
%%CITATION = ARXIV:1302.4028;%%
\bibitem [{\citenamefont {Tscheuschner}(2014)}]{Tscheuschner}%
  \BibitemOpen
  \bibfield  {author} {\bibinfo {author} {\bibfnamefont {J.}~\bibnamefont
  {Tscheuschner}},\ }\emph {\bibinfo {title} {Search for the (Ksi-p) dibaryon
  with ALICE at the LHC}},\ \href@noop {} {Ph.D. thesis},\ \bibinfo  {school}
  {Technische Universit\"at Darmstadt} (\bibinfo {year} {2014})\BibitemShut
  {NoStop}%
\bibitem [{\citenamefont {Adam}\ \emph {et~al.}(2016)\citenamefont {Adam} \emph
  {et~al.}}]{Adam:2015nca}%
  \BibitemOpen
  \bibfield  {author} {\bibinfo {author} {\bibfnamefont {J.}~\bibnamefont
  {Adam}} \emph {et~al.} (\bibinfo {collaboration} {ALICE}),\ }\href {\doibase
  10.1016/j.physletb.2015.11.048} {\bibfield  {journal} {\bibinfo  {journal}
  {Phys. Lett.}\ }\textbf {\bibinfo {volume} {B752}},\ \bibinfo {pages} {267}
  (\bibinfo {year} {2016})},\ \Eprint {http://arxiv.org/abs/1506.07499}
  {arXiv:1506.07499 [nucl-ex]} \BibitemShut {NoStop}%
%%CITATION = ARXIV:1506.07499;%%
\bibitem [{\citenamefont {Badier}\ \emph {et~al.}(1986)\citenamefont {Badier}
  \emph {et~al.}}]{Badier:1986xz}%
  \BibitemOpen
  \bibfield  {author} {\bibinfo {author} {\bibfnamefont {J.}~\bibnamefont
  {Badier}} \emph {et~al.} (\bibinfo {collaboration} {NA3}),\ }\href {\doibase
  10.1007/BF01559588} {\bibfield  {journal} {\bibinfo  {journal} {Z. Phys.}\
  }\textbf {\bibinfo {volume} {C31}},\ \bibinfo {pages} {21} (\bibinfo {year}
  {1986})}\BibitemShut {NoStop}%
%%CITATION = ZEPYA,C31,21;%%
\bibitem [{\citenamefont {Bernstein}\ \emph {et~al.}(1988)\citenamefont
  {Bernstein}, \citenamefont {Shea}, \citenamefont {Winstein}, \citenamefont
  {Cousins}, \citenamefont {Greenhalgh}, \citenamefont {Schwartz},
  \citenamefont {Bock}, \citenamefont {Hedin},\ and\ \citenamefont
  {Thomson}}]{Bernstein:1988ui}%
  \BibitemOpen
  \bibfield  {author} {\bibinfo {author} {\bibfnamefont {R.~H.}\ \bibnamefont
  {Bernstein}}, \bibinfo {author} {\bibfnamefont {T.~K.}\ \bibnamefont {Shea}},
  \bibinfo {author} {\bibfnamefont {B.}~\bibnamefont {Winstein}}, \bibinfo
  {author} {\bibfnamefont {R.~D.}\ \bibnamefont {Cousins}}, \bibinfo {author}
  {\bibfnamefont {J.~F.}\ \bibnamefont {Greenhalgh}}, \bibinfo {author}
  {\bibfnamefont {M.}~\bibnamefont {Schwartz}}, \bibinfo {author}
  {\bibfnamefont {G.~J.}\ \bibnamefont {Bock}}, \bibinfo {author}
  {\bibfnamefont {D.}~\bibnamefont {Hedin}}, \ and\ \bibinfo {author}
  {\bibfnamefont {G.~B.}\ \bibnamefont {Thomson}},\ }\href {\doibase
  10.1103/PhysRevD.37.3103} {\bibfield  {journal} {\bibinfo  {journal} {Phys.
  Rev.}\ }\textbf {\bibinfo {volume} {D37}},\ \bibinfo {pages} {3103} (\bibinfo
  {year} {1988})}\BibitemShut {NoStop}%
%%CITATION = PHRVA,D37,3103;%%
\bibitem [{\citenamefont {Farrar}\ and\ \citenamefont
  {Zaharijas}(2004)}]{Farrar:2003qy}%
  \BibitemOpen
  \bibfield  {author} {\bibinfo {author} {\bibfnamefont {G.~R.}\ \bibnamefont
  {Farrar}}\ and\ \bibinfo {author} {\bibfnamefont {G.}~\bibnamefont
  {Zaharijas}},\ }\href {\doibase 10.1103/PhysRevD.70.014008} {\bibfield
  {journal} {\bibinfo  {journal} {Phys. Rev.}\ }\textbf {\bibinfo {volume}
  {D70}},\ \bibinfo {pages} {014008} (\bibinfo {year} {2004})},\ \Eprint
  {http://arxiv.org/abs/hep-ph/0308137} {arXiv:hep-ph/0308137 [hep-ph]}
  \BibitemShut {NoStop}%
%%CITATION = HEP-PH/0308137;%%
\bibitem [{\citenamefont {Farrar}(2017{\natexlab{a}})}]{Farrar:2017eqq}%
  \BibitemOpen
  \bibfield  {author} {\bibinfo {author} {\bibfnamefont {G.~R.}\ \bibnamefont
  {Farrar}},\ }\href@noop {} {\  (\bibinfo {year} {2017}{\natexlab{a}})},\
  \Eprint {http://arxiv.org/abs/1708.08951} {arXiv:1708.08951 [hep-ph]}
  \BibitemShut {NoStop}%
%%CITATION = ARXIV:1708.08951;%%
\bibitem [{\citenamefont {Farrar}(2017{\natexlab{b}})}]{Farrar:2017ysn}%
  \BibitemOpen
  \bibfield  {author} {\bibinfo {author} {\bibfnamefont {G.~R.}\ \bibnamefont
  {Farrar}},\ }in\ \href
  {http://inspirehep.net/record/1639411/files/arXiv:1711.10971.pdf} {\emph
  {\bibinfo {booktitle} {{Proceedings, 35th International Cosmic Ray Conference
  (ICRC 2017): Bexco, Busan, Korea, July 12-20, 2017}}}}\ (\bibinfo {year}
  {2017})\ \Eprint {http://arxiv.org/abs/1711.10971} {arXiv:1711.10971
  [hep-ph]} \BibitemShut {NoStop}%
%%CITATION = ARXIV:1711.10971;%%
\bibitem [{\citenamefont {Zaharijas}\ and\ \citenamefont
  {Farrar}(2005)}]{Zaharijas:2004jv}%
  \BibitemOpen
  \bibfield  {author} {\bibinfo {author} {\bibfnamefont {G.}~\bibnamefont
  {Zaharijas}}\ and\ \bibinfo {author} {\bibfnamefont {G.~R.}\ \bibnamefont
  {Farrar}},\ }\href {\doibase 10.1103/PhysRevD.72.083502} {\bibfield
  {journal} {\bibinfo  {journal} {Phys. Rev.}\ }\textbf {\bibinfo {volume}
  {D72}},\ \bibinfo {pages} {083502} (\bibinfo {year} {2005})},\ \Eprint
  {http://arxiv.org/abs/astro-ph/0406531} {arXiv:astro-ph/0406531 [astro-ph]}
  \BibitemShut {NoStop}%
%%CITATION = ASTRO-PH/0406531;%%
\bibitem [{\citenamefont {Farrar}(2018)}]{Farrar:2018hac}%
  \BibitemOpen
  \bibfield  {author} {\bibinfo {author} {\bibfnamefont {G.~R.}\ \bibnamefont
  {Farrar}},\ }\href@noop {} {\  (\bibinfo {year} {2018})},\ \Eprint
  {http://arxiv.org/abs/1805.03723} {arXiv:1805.03723 [hep-ph]} \BibitemShut
  {NoStop}%
%%CITATION = ARXIV:1805.03723;%%
\bibitem [{\citenamefont {Beane}\ \emph
  {et~al.}(2011{\natexlab{a}})\citenamefont {Beane} \emph
  {et~al.}}]{Beane:2010hg}%
  \BibitemOpen
  \bibfield  {author} {\bibinfo {author} {\bibfnamefont {S.~R.}\ \bibnamefont
  {Beane}} \emph {et~al.} (\bibinfo {collaboration} {NPLQCD}),\ }\href
  {\doibase 10.1103/PhysRevLett.106.162001} {\bibfield  {journal} {\bibinfo
  {journal} {Phys. Rev. Lett.}\ }\textbf {\bibinfo {volume} {106}},\ \bibinfo
  {pages} {162001} (\bibinfo {year} {2011}{\natexlab{a}})},\ \Eprint
  {http://arxiv.org/abs/1012.3812} {arXiv:1012.3812 [hep-lat]} \BibitemShut
  {NoStop}%
%%CITATION = ARXIV:1012.3812;%%
\bibitem [{\citenamefont {Beane}\ \emph
  {et~al.}(2011{\natexlab{b}})\citenamefont {Beane} \emph
  {et~al.}}]{Beane:2011zpa}%
  \BibitemOpen
  \bibfield  {author} {\bibinfo {author} {\bibfnamefont {S.~R.}\ \bibnamefont
  {Beane}} \emph {et~al.},\ }\href {\doibase 10.1142/S0217732311036978}
  {\bibfield  {journal} {\bibinfo  {journal} {Mod. Phys. Lett.}\ }\textbf
  {\bibinfo {volume} {A26}},\ \bibinfo {pages} {2587} (\bibinfo {year}
  {2011}{\natexlab{b}})},\ \Eprint {http://arxiv.org/abs/1103.2821}
  {arXiv:1103.2821 [hep-lat]} \BibitemShut {NoStop}%
%%CITATION = ARXIV:1103.2821;%%
\bibitem [{\citenamefont {Beane}\ \emph {et~al.}(2012)\citenamefont {Beane},
  \citenamefont {Chang}, \citenamefont {Detmold}, \citenamefont {Lin},
  \citenamefont {Luu}, \citenamefont {Orginos}, \citenamefont {Parreno},
  \citenamefont {Savage}, \citenamefont {Torok},\ and\ \citenamefont
  {Walker-Loud}}]{Beane:2011iw}%
  \BibitemOpen
  \bibfield  {author} {\bibinfo {author} {\bibfnamefont {S.~R.}\ \bibnamefont
  {Beane}}, \bibinfo {author} {\bibfnamefont {E.}~\bibnamefont {Chang}},
  \bibinfo {author} {\bibfnamefont {W.}~\bibnamefont {Detmold}}, \bibinfo
  {author} {\bibfnamefont {H.~W.}\ \bibnamefont {Lin}}, \bibinfo {author}
  {\bibfnamefont {T.~C.}\ \bibnamefont {Luu}}, \bibinfo {author} {\bibfnamefont
  {K.}~\bibnamefont {Orginos}}, \bibinfo {author} {\bibfnamefont
  {A.}~\bibnamefont {Parreno}}, \bibinfo {author} {\bibfnamefont {M.~J.}\
  \bibnamefont {Savage}}, \bibinfo {author} {\bibfnamefont {A.}~\bibnamefont
  {Torok}}, \ and\ \bibinfo {author} {\bibfnamefont {A.}~\bibnamefont
  {Walker-Loud}} (\bibinfo {collaboration} {NPLQCD}),\ }\href {\doibase
  10.1103/PhysRevD.85.054511} {\bibfield  {journal} {\bibinfo  {journal} {Phys.
  Rev.}\ }\textbf {\bibinfo {volume} {D85}},\ \bibinfo {pages} {054511}
  (\bibinfo {year} {2012})},\ \Eprint {http://arxiv.org/abs/1109.2889}
  {arXiv:1109.2889 [hep-lat]} \BibitemShut {NoStop}%
%%CITATION = ARXIV:1109.2889;%%
\bibitem [{\citenamefont {Beane}\ \emph {et~al.}(2013)\citenamefont {Beane},
  \citenamefont {Chang}, \citenamefont {Cohen}, \citenamefont {Detmold},
  \citenamefont {Lin}, \citenamefont {Luu}, \citenamefont {Orginos},
  \citenamefont {Parreno}, \citenamefont {Savage},\ and\ \citenamefont
  {Walker-Loud}}]{Beane:2012vq}%
  \BibitemOpen
  \bibfield  {author} {\bibinfo {author} {\bibfnamefont {S.~R.}\ \bibnamefont
  {Beane}}, \bibinfo {author} {\bibfnamefont {E.}~\bibnamefont {Chang}},
  \bibinfo {author} {\bibfnamefont {S.~D.}\ \bibnamefont {Cohen}}, \bibinfo
  {author} {\bibfnamefont {W.}~\bibnamefont {Detmold}}, \bibinfo {author}
  {\bibfnamefont {H.~W.}\ \bibnamefont {Lin}}, \bibinfo {author} {\bibfnamefont
  {T.~C.}\ \bibnamefont {Luu}}, \bibinfo {author} {\bibfnamefont
  {K.}~\bibnamefont {Orginos}}, \bibinfo {author} {\bibfnamefont
  {A.}~\bibnamefont {Parreno}}, \bibinfo {author} {\bibfnamefont {M.~J.}\
  \bibnamefont {Savage}}, \ and\ \bibinfo {author} {\bibfnamefont
  {A.}~\bibnamefont {Walker-Loud}} (\bibinfo {collaboration} {NPLQCD}),\ }\href
  {\doibase 10.1103/PhysRevD.87.034506} {\bibfield  {journal} {\bibinfo
  {journal} {Phys. Rev.}\ }\textbf {\bibinfo {volume} {D87}},\ \bibinfo {pages}
  {034506} (\bibinfo {year} {2013})},\ \Eprint {http://arxiv.org/abs/1206.5219}
  {arXiv:1206.5219 [hep-lat]} \BibitemShut {NoStop}%
%%CITATION = ARXIV:1206.5219;%%
\bibitem [{\citenamefont {Savage}\ and\ \citenamefont
  {Beane}()}]{SavageBeaneCommunication}%
  \BibitemOpen
  \bibfield  {author} {\bibinfo {author} {\bibfnamefont {M.~J.}\ \bibnamefont
  {Savage}}\ and\ \bibinfo {author} {\bibfnamefont {S.~R.}\ \bibnamefont
  {Beane}},\ }\href@noop {} {\emph {\bibinfo {title} {private
  communication}}}\BibitemShut {NoStop}%
\bibitem [{\citenamefont {Mahdawi}\ and\ \citenamefont
  {Farrar}(2018)}]{Mahdawi:2018euy}%
  \BibitemOpen
  \bibfield  {author} {\bibinfo {author} {\bibfnamefont {M.~S.}\ \bibnamefont
  {Mahdawi}}\ and\ \bibinfo {author} {\bibfnamefont {G.~R.}\ \bibnamefont
  {Farrar}},\ }\href {\doibase 10.1088/1475-7516/2018/10/007} {\bibfield
  {journal} {\bibinfo  {journal} {JCAP}\ }\textbf {\bibinfo {volume} {1810}},\
  \bibinfo {pages} {007} (\bibinfo {year} {2018})},\ \Eprint
  {http://arxiv.org/abs/1804.03073} {arXiv:1804.03073 [hep-ph]} \BibitemShut
  {NoStop}%
%%CITATION = ARXIV:1804.03073;%%
\bibitem [{\citenamefont {Gluscevic}\ and\ \citenamefont
  {Boddy}(2018)}]{Gluscevic:2017ywp}%
  \BibitemOpen
  \bibfield  {author} {\bibinfo {author} {\bibfnamefont {V.}~\bibnamefont
  {Gluscevic}}\ and\ \bibinfo {author} {\bibfnamefont {K.~K.}\ \bibnamefont
  {Boddy}},\ }\href {\doibase 10.1103/PhysRevLett.121.081301} {\bibfield
  {journal} {\bibinfo  {journal} {Phys. Rev. Lett.}\ }\textbf {\bibinfo
  {volume} {121}},\ \bibinfo {pages} {081301} (\bibinfo {year} {2018})},\
  \Eprint {http://arxiv.org/abs/1712.07133} {arXiv:1712.07133 [astro-ph.CO]}
  \BibitemShut {NoStop}%
%%CITATION = ARXIV:1712.07133;%%
\bibitem [{\citenamefont {Hooper}\ and\ \citenamefont
  {McDermott}(2018)}]{Hooper:2018bfw}%
  \BibitemOpen
  \bibfield  {author} {\bibinfo {author} {\bibfnamefont {D.}~\bibnamefont
  {Hooper}}\ and\ \bibinfo {author} {\bibfnamefont {S.~D.}\ \bibnamefont
  {McDermott}},\ }\href {\doibase 10.1103/PhysRevD.97.115006} {\bibfield
  {journal} {\bibinfo  {journal} {Phys. Rev.}\ }\textbf {\bibinfo {volume}
  {D97}},\ \bibinfo {pages} {115006} (\bibinfo {year} {2018})},\ \Eprint
  {http://arxiv.org/abs/1802.03025} {arXiv:1802.03025 [hep-ph]} \BibitemShut
  {NoStop}%
%%CITATION = ARXIV:1802.03025;%%
\bibitem [{\citenamefont {Pons}\ \emph {et~al.}(1999)\citenamefont {Pons},
  \citenamefont {Reddy}, \citenamefont {Prakash}, \citenamefont {Lattimer},\
  and\ \citenamefont {Miralles}}]{Pons:1998mm}%
  \BibitemOpen
  \bibfield  {author} {\bibinfo {author} {\bibfnamefont {J.~A.}\ \bibnamefont
  {Pons}}, \bibinfo {author} {\bibfnamefont {S.}~\bibnamefont {Reddy}},
  \bibinfo {author} {\bibfnamefont {M.}~\bibnamefont {Prakash}}, \bibinfo
  {author} {\bibfnamefont {J.~M.}\ \bibnamefont {Lattimer}}, \ and\ \bibinfo
  {author} {\bibfnamefont {J.~A.}\ \bibnamefont {Miralles}},\ }\href {\doibase
  10.1086/306889} {\bibfield  {journal} {\bibinfo  {journal} {Astrophys. J.}\
  }\textbf {\bibinfo {volume} {513}},\ \bibinfo {pages} {780} (\bibinfo {year}
  {1999})},\ \Eprint {http://arxiv.org/abs/astro-ph/9807040}
  {arXiv:astro-ph/9807040 [astro-ph]} \BibitemShut {NoStop}%
%%CITATION = ASTRO-PH/9807040;%%
\bibitem [{\citenamefont {Keil}\ and\ \citenamefont
  {Janka}(1995)}]{Keil:1995hw}%
  \BibitemOpen
  \bibfield  {author} {\bibinfo {author} {\bibfnamefont {W.}~\bibnamefont
  {Keil}}\ and\ \bibinfo {author} {\bibfnamefont {H.~T.}\ \bibnamefont
  {Janka}},\ }\href@noop {} {\bibfield  {journal} {\bibinfo  {journal} {Astron.
  Astrophys.}\ }\textbf {\bibinfo {volume} {296}},\ \bibinfo {pages} {145}
  (\bibinfo {year} {1995})}\BibitemShut {NoStop}%
%%CITATION = AAEJA,296,145;%%
\bibitem [{\citenamefont {Burrows}\ and\ \citenamefont
  {Lattimer}(1986)}]{Burrows:1986me}%
  \BibitemOpen
  \bibfield  {author} {\bibinfo {author} {\bibfnamefont {A.}~\bibnamefont
  {Burrows}}\ and\ \bibinfo {author} {\bibfnamefont {J.~M.}\ \bibnamefont
  {Lattimer}},\ }\href {\doibase 10.1086/164405} {\bibfield  {journal}
  {\bibinfo  {journal} {Astrophys. J.}\ }\textbf {\bibinfo {volume} {307}},\
  \bibinfo {pages} {178} (\bibinfo {year} {1986})}\BibitemShut {NoStop}%
%%CITATION = ASJOA,307,178;%%
\bibitem [{Note1()}]{Note1}%
  \BibitemOpen
  \bibinfo {note} {We thank the authors of \cite {KolbTurnerdraft} for pointing
  out that strong isospin forbids $\Lambda \Lambda \to S \pi $.}\BibitemShut
  {Stop}%
\bibitem [{\citenamefont {Brinkmann}\ and\ \citenamefont
  {Turner}(1988)}]{Brinkmann:1988vi}%
  \BibitemOpen
  \bibfield  {author} {\bibinfo {author} {\bibfnamefont {R.~P.}\ \bibnamefont
  {Brinkmann}}\ and\ \bibinfo {author} {\bibfnamefont {M.~S.}\ \bibnamefont
  {Turner}},\ }\href {\doibase 10.1103/PhysRevD.38.2338} {\bibfield  {journal}
  {\bibinfo  {journal} {Phys. Rev.}\ }\textbf {\bibinfo {volume} {D38}},\
  \bibinfo {pages} {2338} (\bibinfo {year} {1988})}\BibitemShut {NoStop}%
%%CITATION = PHRVA,D38,2338;%%
\bibitem [{\citenamefont {Raffelt}\ and\ \citenamefont
  {Seckel}(1995)}]{Raffelt:1993ix}%
  \BibitemOpen
  \bibfield  {author} {\bibinfo {author} {\bibfnamefont {G.}~\bibnamefont
  {Raffelt}}\ and\ \bibinfo {author} {\bibfnamefont {D.}~\bibnamefont
  {Seckel}},\ }\href {\doibase 10.1103/PhysRevD.52.1780} {\bibfield  {journal}
  {\bibinfo  {journal} {Phys. Rev.}\ }\textbf {\bibinfo {volume} {D52}},\
  \bibinfo {pages} {1780} (\bibinfo {year} {1995})},\ \Eprint
  {http://arxiv.org/abs/astro-ph/9312019} {arXiv:astro-ph/9312019 [astro-ph]}
  \BibitemShut {NoStop}%
%%CITATION = ASTRO-PH/9312019;%%
\bibitem [{\citenamefont {Raffelt}(2008)}]{Raffelt:2006cw}%
  \BibitemOpen
  \bibfield  {author} {\bibinfo {author} {\bibfnamefont {G.~G.}\ \bibnamefont
  {Raffelt}},\ }\bibfield  {booktitle} {\emph {\bibinfo {booktitle} {{Axions:
  Theory, cosmology, and experimental searches. Proceedings, 1st Joint
  ILIAS-CERN-CAST axion training, Geneva, Switzerland, November 30-December 2,
  2005}}},\ }\href {\doibase 10.1007/978-3-540-73518-2_3} {\bibfield  {journal}
  {\bibinfo  {journal} {Lect. Notes Phys.}\ }\textbf {\bibinfo {volume}
  {741}},\ \bibinfo {pages} {51} (\bibinfo {year} {2008})},\ \bibinfo {note}
  {[,51(2006)]},\ \Eprint {http://arxiv.org/abs/hep-ph/0611350}
  {arXiv:hep-ph/0611350 [hep-ph]} \BibitemShut {NoStop}%
%%CITATION = HEP-PH/0611350;%%
\bibitem [{\citenamefont {Gross}\ \emph {et~al.}(2018)\citenamefont {Gross},
  \citenamefont {Polosa}, \citenamefont {Strumia}, \citenamefont {Urbano},\
  and\ \citenamefont {Xue}}]{Gross:2018ivp}%
  \BibitemOpen
  \bibfield  {author} {\bibinfo {author} {\bibfnamefont {C.}~\bibnamefont
  {Gross}}, \bibinfo {author} {\bibfnamefont {A.}~\bibnamefont {Polosa}},
  \bibinfo {author} {\bibfnamefont {A.}~\bibnamefont {Strumia}}, \bibinfo
  {author} {\bibfnamefont {A.}~\bibnamefont {Urbano}}, \ and\ \bibinfo {author}
  {\bibfnamefont {W.}~\bibnamefont {Xue}},\ }\href@noop {} {\  (\bibinfo {year}
  {2018})},\ \Eprint {http://arxiv.org/abs/1803.10242} {arXiv:1803.10242
  [hep-ph]} \BibitemShut {NoStop}%
%%CITATION = ARXIV:1803.10242;%%
\bibitem [{\citenamefont {Isgur}\ and\ \citenamefont
  {Karl}(1979)}]{Isgur:1978wd}%
  \BibitemOpen
  \bibfield  {author} {\bibinfo {author} {\bibfnamefont {N.}~\bibnamefont
  {Isgur}}\ and\ \bibinfo {author} {\bibfnamefont {G.}~\bibnamefont {Karl}},\
  }\href {\doibase 10.1103/PhysRevD.23.817.2, 10.1103/PhysRevD.19.2653}
  {\bibfield  {journal} {\bibinfo  {journal} {Phys. Rev.}\ }\textbf {\bibinfo
  {volume} {D19}},\ \bibinfo {pages} {2653} (\bibinfo {year} {1979})},\
  \bibinfo {note} {[Erratum: Phys. Rev.D23,817(1981)]}\BibitemShut {NoStop}%
%%CITATION = PHRVA,D19,2653;%%
\bibitem [{\citenamefont {Day}(1967)}]{DAY:1967aa}%
  \BibitemOpen
  \bibfield  {author} {\bibinfo {author} {\bibfnamefont {B.~D.}\ \bibnamefont
  {Day}},\ }\href {\doibase 10.1103/RevModPhys.39.719} {\bibfield  {journal}
  {\bibinfo  {journal} {Reviews of Modern Physics}\ }\textbf {\bibinfo {volume}
  {39}},\ \bibinfo {pages} {719} (\bibinfo {year} {1967})}\BibitemShut
  {NoStop}%
\bibitem [{\citenamefont {Faessler}\ \emph {et~al.}(1997)\citenamefont
  {Faessler}, \citenamefont {Buchmann},\ and\ \citenamefont
  {Krivoruchenko}}]{Faessler:1997jg}%
  \BibitemOpen
  \bibfield  {author} {\bibinfo {author} {\bibfnamefont {A.}~\bibnamefont
  {Faessler}}, \bibinfo {author} {\bibfnamefont {A.~J.}\ \bibnamefont
  {Buchmann}}, \ and\ \bibinfo {author} {\bibfnamefont {M.~I.}\ \bibnamefont
  {Krivoruchenko}},\ }\href {\doibase 10.1103/PhysRevC.56.1576} {\bibfield
  {journal} {\bibinfo  {journal} {Phys. Rev.}\ }\textbf {\bibinfo {volume}
  {C56}},\ \bibinfo {pages} {1576} (\bibinfo {year} {1997})},\ \Eprint
  {http://arxiv.org/abs/nucl-th/9706080} {arXiv:nucl-th/9706080 [nucl-th]}
  \BibitemShut {NoStop}%
%%CITATION = NUCL-TH/9706080;%%
\bibitem [{\citenamefont {Povh}\ and\ \citenamefont
  {Hufner}(1990)}]{Povh:1990ad}%
  \BibitemOpen
  \bibfield  {author} {\bibinfo {author} {\bibfnamefont {B.}~\bibnamefont
  {Povh}}\ and\ \bibinfo {author} {\bibfnamefont {J.}~\bibnamefont {Hufner}},\
  }\href {\doibase 10.1016/0370-2693(90)90707-D} {\bibfield  {journal}
  {\bibinfo  {journal} {Phys. Lett.}\ }\textbf {\bibinfo {volume} {B245}},\
  \bibinfo {pages} {653} (\bibinfo {year} {1990})}\BibitemShut {NoStop}%
%%CITATION = PHLTA,B245,653;%%
\bibitem [{\citenamefont {Demorest}\ \emph {et~al.}(2010)\citenamefont
  {Demorest}, \citenamefont {Pennucci}, \citenamefont {Ransom}, \citenamefont
  {Roberts},\ and\ \citenamefont {Hessels}}]{Demorest:2010bx}%
  \BibitemOpen
  \bibfield  {author} {\bibinfo {author} {\bibfnamefont {P.}~\bibnamefont
  {Demorest}}, \bibinfo {author} {\bibfnamefont {T.}~\bibnamefont {Pennucci}},
  \bibinfo {author} {\bibfnamefont {S.}~\bibnamefont {Ransom}}, \bibinfo
  {author} {\bibfnamefont {M.}~\bibnamefont {Roberts}}, \ and\ \bibinfo
  {author} {\bibfnamefont {J.}~\bibnamefont {Hessels}},\ }\href {\doibase
  10.1038/nature09466} {\bibfield  {journal} {\bibinfo  {journal} {Nature}\
  }\textbf {\bibinfo {volume} {467}},\ \bibinfo {pages} {1081} (\bibinfo {year}
  {2010})},\ \Eprint {http://arxiv.org/abs/1010.5788} {arXiv:1010.5788
  [astro-ph.HE]} \BibitemShut {NoStop}%
%%CITATION = ARXIV:1010.5788;%%
\bibitem [{\citenamefont {Antoniadis}\ \emph {et~al.}(2013)\citenamefont
  {Antoniadis} \emph {et~al.}}]{Antoniadis:2013pzd}%
  \BibitemOpen
  \bibfield  {author} {\bibinfo {author} {\bibfnamefont {J.}~\bibnamefont
  {Antoniadis}} \emph {et~al.},\ }\href {\doibase 10.1126/science.1233232}
  {\bibfield  {journal} {\bibinfo  {journal} {Science}\ }\textbf {\bibinfo
  {volume} {340}},\ \bibinfo {pages} {6131} (\bibinfo {year} {2013})},\ \Eprint
  {http://arxiv.org/abs/1304.6875} {arXiv:1304.6875 [astro-ph.HE]} \BibitemShut
  {NoStop}%
%%CITATION = ARXIV:1304.6875;%%
\bibitem [{\citenamefont {Bauswein}\ \emph {et~al.}(2017)\citenamefont
  {Bauswein}, \citenamefont {Just}, \citenamefont {Janka},\ and\ \citenamefont
  {Stergioulas}}]{Bauswein:2017vtn}%
  \BibitemOpen
  \bibfield  {author} {\bibinfo {author} {\bibfnamefont {A.}~\bibnamefont
  {Bauswein}}, \bibinfo {author} {\bibfnamefont {O.}~\bibnamefont {Just}},
  \bibinfo {author} {\bibfnamefont {H.-T.}\ \bibnamefont {Janka}}, \ and\
  \bibinfo {author} {\bibfnamefont {N.}~\bibnamefont {Stergioulas}},\ }\href
  {\doibase 10.3847/2041-8213/aa9994} {\bibfield  {journal} {\bibinfo
  {journal} {Astrophys. J.}\ }\textbf {\bibinfo {volume} {850}},\ \bibinfo
  {pages} {L34} (\bibinfo {year} {2017})},\ \Eprint
  {http://arxiv.org/abs/1710.06843} {arXiv:1710.06843 [astro-ph.HE]}
  \BibitemShut {NoStop}%
%%CITATION = ARXIV:1710.06843;%%
\bibitem [{\citenamefont {Annala}\ \emph {et~al.}(2018)\citenamefont {Annala},
  \citenamefont {Gorda}, \citenamefont {Kurkela},\ and\ \citenamefont
  {Vuorinen}}]{Annala:2017llu}%
  \BibitemOpen
  \bibfield  {author} {\bibinfo {author} {\bibfnamefont {E.}~\bibnamefont
  {Annala}}, \bibinfo {author} {\bibfnamefont {T.}~\bibnamefont {Gorda}},
  \bibinfo {author} {\bibfnamefont {A.}~\bibnamefont {Kurkela}}, \ and\
  \bibinfo {author} {\bibfnamefont {A.}~\bibnamefont {Vuorinen}},\ }\href
  {\doibase 10.1103/PhysRevLett.120.172703} {\bibfield  {journal} {\bibinfo
  {journal} {Phys. Rev. Lett.}\ }\textbf {\bibinfo {volume} {120}},\ \bibinfo
  {pages} {172703} (\bibinfo {year} {2018})},\ \Eprint
  {http://arxiv.org/abs/1711.02644} {arXiv:1711.02644 [astro-ph.HE]}
  \BibitemShut {NoStop}%
%%CITATION = ARXIV:1711.02644;%%
\bibitem [{\citenamefont {Most}\ \emph {et~al.}(2018)\citenamefont {Most},
  \citenamefont {Weih}, \citenamefont {Rezzolla},\ and\ \citenamefont
  {Schaffner-Bielich}}]{Most:2018hfd}%
  \BibitemOpen
  \bibfield  {author} {\bibinfo {author} {\bibfnamefont {E.~R.}\ \bibnamefont
  {Most}}, \bibinfo {author} {\bibfnamefont {L.~R.}\ \bibnamefont {Weih}},
  \bibinfo {author} {\bibfnamefont {L.}~\bibnamefont {Rezzolla}}, \ and\
  \bibinfo {author} {\bibfnamefont {J.}~\bibnamefont {Schaffner-Bielich}},\
  }\href {\doibase 10.1103/PhysRevLett.120.261103} {\bibfield  {journal}
  {\bibinfo  {journal} {Phys. Rev. Lett.}\ }\textbf {\bibinfo {volume} {120}},\
  \bibinfo {pages} {261103} (\bibinfo {year} {2018})},\ \Eprint
  {http://arxiv.org/abs/1803.00549} {arXiv:1803.00549 [gr-qc]} \BibitemShut
  {NoStop}%
%%CITATION = ARXIV:1803.00549;%%
\bibitem [{\citenamefont {Kolb}\ and\ \citenamefont
  {Turner}(2018)}]{KolbTurnerdraft}%
  \BibitemOpen
  \bibfield  {author} {\bibinfo {author} {\bibfnamefont {E.~W.}\ \bibnamefont
  {Kolb}}\ and\ \bibinfo {author} {\bibfnamefont {M.}~\bibnamefont {Turner}},\
  }\href@noop {} {\  (\bibinfo {year} {2018})},\ \Eprint
  {http://arxiv.org/abs/1809.06003} {arXiv:1809.06003 [hep-ph]} \BibitemShut
  {NoStop}%
%%CITATION = ARXIV: 1809.06003;%%
\bibitem [{\citenamefont {Lees}\ \emph {et~al.}(2018)\citenamefont {Lees} \emph
  {et~al.}}]{Echenard:2018rfe}%
  \BibitemOpen
  \bibfield  {author} {\bibinfo {author} {\bibfnamefont {J.~P.}\ \bibnamefont
  {Lees}} \emph {et~al.} (\bibinfo {collaboration} {BaBar}),\ }\href@noop {} {\
   (\bibinfo {year} {2018})},\ \Eprint {http://arxiv.org/abs/1810.04724}
  {arXiv:1810.04724 [hep-ex]} \BibitemShut {NoStop}%
%%CITATION = ARXIV:1810.04724;%%
\bibitem [{\citenamefont {Morita}\ \emph {et~al.}(2015)\citenamefont {Morita},
  \citenamefont {Furumoto},\ and\ \citenamefont {Ohnishi}}]{Morita:2014kza}%
  \BibitemOpen
  \bibfield  {author} {\bibinfo {author} {\bibfnamefont {K.}~\bibnamefont
  {Morita}}, \bibinfo {author} {\bibfnamefont {T.}~\bibnamefont {Furumoto}}, \
  and\ \bibinfo {author} {\bibfnamefont {A.}~\bibnamefont {Ohnishi}},\ }\href
  {\doibase 10.1103/PhysRevC.91.024916} {\bibfield  {journal} {\bibinfo
  {journal} {Phys. Rev.}\ }\textbf {\bibinfo {volume} {C91}},\ \bibinfo {pages}
  {024916} (\bibinfo {year} {2015})},\ \Eprint {http://arxiv.org/abs/1408.6682}
  {arXiv:1408.6682 [nucl-th]} \BibitemShut {NoStop}%
%%CITATION = ARXIV:1408.6682;%%
\bibitem [{\citenamefont {Haidenbauer}\ \emph {et~al.}(2016)\citenamefont
  {Haidenbauer}, \citenamefont {Meiner},\ and\ \citenamefont
  {Petschauer}}]{Haidenbauer:2015zqb}%
  \BibitemOpen
  \bibfield  {author} {\bibinfo {author} {\bibfnamefont {J.}~\bibnamefont
  {Haidenbauer}}, \bibinfo {author} {\bibfnamefont {U.-G.}\ \bibnamefont
  {Meiner}}, \ and\ \bibinfo {author} {\bibfnamefont {S.}~\bibnamefont
  {Petschauer}},\ }\href {\doibase 10.1016/j.nuclphysa.2016.01.006} {\bibfield
  {journal} {\bibinfo  {journal} {Nucl. Phys.}\ }\textbf {\bibinfo {volume}
  {A954}},\ \bibinfo {pages} {273} (\bibinfo {year} {2016})},\ \Eprint
  {http://arxiv.org/abs/1511.05859} {arXiv:1511.05859 [nucl-th]} \BibitemShut
  {NoStop}%
%%CITATION = ARXIV:1511.05859;%%
\bibitem [{\citenamefont {Li}\ \emph {et~al.}(2018)\citenamefont {Li},
  \citenamefont {Hyodo},\ and\ \citenamefont {Geng}}]{Li:2018tbt}%
  \BibitemOpen
  \bibfield  {author} {\bibinfo {author} {\bibfnamefont {K.-W.}\ \bibnamefont
  {Li}}, \bibinfo {author} {\bibfnamefont {T.}~\bibnamefont {Hyodo}}, \ and\
  \bibinfo {author} {\bibfnamefont {L.-S.}\ \bibnamefont {Geng}},\ }\href@noop
  {} {\  (\bibinfo {year} {2018})},\ \Eprint {http://arxiv.org/abs/1809.03199}
  {arXiv:1809.03199 [nucl-th]} \BibitemShut {NoStop}%
%%CITATION = ARXIV:1809.03199;%%
\end{thebibliography}%

\end{document}